\documentclass[a4paper,11pt]{article}
\pdfoutput=1 

\usepackage{jcappub}

\usepackage[T1]{fontenc} 
\usepackage{graphicx}	
\usepackage{amsmath}	
\usepackage{amssymb}	
\usepackage{mathtools}
\usepackage{verbatim}
\usepackage{natbib}
\usepackage{color}
\definecolor{darkblue}{rgb}{0,0.2,0.8}
\definecolor{darkred}{rgb}{0.75,0,0}
\def\simgt{\hbox{\,\rlap{\raise 0.425ex\hbox{$>$}}\lower 0.65ex\hbox{$\sim$}\,}}
\def\simlt{\hbox{\,\rlap{\raise 0.425ex\hbox{$<$}}\lower 0.65ex\hbox{$\sim$}\,}}

\title{\boldmath Ubiquity of density slope oscillations in the central regions of galaxy and
cluster-sized systems}

\author[1]{Anthony M. Young}
\author[1]{, Liliya L. R. Williams}
\author[2]{, and Jens Hjorth }

\affiliation[1]{School of Physics and Astronomy, University of Minnesota, 116 Church Street SE, Minneapolis, MN 55455, USA}
\affiliation[2]{Dark Cosmology Centre, Niels Bohr Institute, University of Copenhagen, Juliane Maries Vej 30, DK-2100 Copenhagen, Denmark}

\emailAdd{amyoung@astro.umn.edu,llrw@astro.umn.edu,jens@dark-cosmology.dk}

\abstract{
One usually thinks of a radial density profile as having a monotonically changing
logarithmic slope, such as in NFW or Einasto profiles. However, in two different
classes of commonly used systems, this is often not the case. These classes exhibit
non-monotonic changes in their density profile slopes which we call oscillations for
short.  We analyze these two unrelated classes separately. Class 1 consists of
systems that have density oscillations and that are defined through their
distribution function $f(E)$, or differential energy distribution $N(E)$, such as
isothermal spheres, King profiles, or DARKexp, a theoretically derived model for
relaxed collisionless systems. Systems defined through $f(E)$ or $N(E)$ generally
have density slope oscillations. Class 1 system oscillations can be found at small,
intermediate, or large radii but we focus on a limited set of Class 1 systems that
have oscillations in the central regions, usually at $\log(r/r_{-2})\lesssim -2$,
where $r_{-2}$ is the largest radius where $d\log(\rho)/d\log(r)=-2$. We show that
the shape of their $N(E)$ can roughly predict the amplitude of oscillations. Class 2 
systems which are a product of dynamical evolution, consist of observed and simulated 
galaxies and clusters, and pure dark matter halos. Oscillations in the density profile 
slope seem pervasive in the central regions of Class 2 systems. We argue that in these 
systems, slope oscillations are an indication that a system is not fully relaxed.
We show that these oscillations can be reproduced by small modifications to $N(E)$
of DARKexp. These affect a small fraction of systems' mass and are confined to
$\log(r/r_{-2})\lesssim 0$. The size of these modifications serves as a potential
diagnostic for quantifying how far a system is from being relaxed.
} 

\begin{document}
\maketitle
\flushbottom

\section{Introduction}

The distribution of dark matter in galaxies and clusters of galaxies is well described by $\Lambda$CDM numerical N-body simulations of cosmological 
structure formation and evolution.  These simulations produce near universal dark matter halo radial density profiles over a few decades in radius 
\citep{Nav04, Sta09}. In general, the Navarro-Frenk-White (NFW) profile \citep{Nav97} provides a good fit to density profiles of equilibrium dark matter 
halos \citep{Tas04, Die04, Die05a, Die05b, Nav10, Die11, Lud11}.  However, there are some discrepancies between simulations and 
observations.  For example, the core/cusp problem \citep{Flo94}: N-Body simulations produce cuspy central density profiles \citep{Dub99}, well fit by the NFW 
profile, while some observations show core like centers in dark matter dominated systems, like dwarf galaxies \citep{Pon14, Wei15}.

\begin{table}
\label{Models}
    \begin{tabular}{| ll |}
    \hline
    \bf{Class 1 Systems}&\\ \hline
DARKexp profiles \citep{Hjo10}& $N(\epsilon)=A[e^{\phi_0-\epsilon}-1]$\\
Polytrope \citep{Med01, Fer08}& $f(\epsilon)=A\epsilon^{n-3/2}, \quad n=5\pm\eta\quad(\eta<<1), \quad n>5$\\
Isothermal Sphere (IS) \citep{Bin87}& $f(\epsilon)=\frac{\rho_0}{(2\pi\sigma^2)^{3/2}}e^{\epsilon/\sigma^2}$\\
Lowered Isothermal Models \citep{Gie15}& $f(E,J^2)=A\exp\left(-\dfrac{J^2}{2r^2_as^2}\right)E_\gamma \left(g,-\dfrac{E-\phi(r_t)}{s^2}\right)$\\
King profiles \citep{Kin66}& $f(\epsilon)= \rho_1(2\pi\sigma^2)^{-3/2}(\exp(\epsilon/\sigma^2)-1)$\\
    \hline
\bf{Class 2 Systems}&\\ \hline
SDSS elliptical galaxies \citep{Cha14}&\\
Massive galaxy clusters \citep{New13}&\\
GHALO$_2^1$ ($\epsilon=61.0 pc$) \citep{Sta09}&\\
Via Lactea 2$^1$($\epsilon=40.0 pc$) \citep{Die08}&\\
EAGLE$^2$ ($\epsilon=700 pc$) \citep{Schay15}&\\
    \hline
\bf{Non Oscillating Systems}&\\ \hline
Hernquist profiles \citep{Her90}& $\rho(r)=\frac{M}{2\pi}\frac{a}{r}\frac{1}{(r+a)^3}$\\
Polytrope \citep{Med01, Fer08}& $\rho(r)=\rho_c\Psi^n$ e.g. $  \Psi_{0} = 1-r^2/6, \quad  \Psi_{1} = \dfrac{\sin r}{r},\quad  \Psi_{5} = (1+\dfrac{1}{3}r^2)^{-1/2}$\\
NFW and gNFW profiles \citep{Nav97}& $\rho(r) = \dfrac{\rho_s}{(r/r_s)^\alpha(1+r/r_s)^{3-\alpha}}$ \\
Einasto profiles \citep{Ein65}& $\rho(r)=\rho_{s}\exp \{-b_n[(r/r_{s})^{1/n}-1]\}$\\
\citet{Nav04} profiles$^{1,2}$&\\
Millennium-II$^2$ ($\epsilon=1370 pc$) \citep{Boy09}& \\
Aquarius$^2$(AQ-1 $\epsilon=20.5 pc$) \citep{Spr08}& \\
    \hline
\end{tabular}
\caption{ An incomplete list of density profiles. Class 1 systems consist of systems defined through a physics based approach that have density slope 
oscillations.  Class 2 systems are dynamically evolved systems with density slope oscillations.  Note that polytropic models 
can produce oscillations or no oscillations in density slope depending on the polytropic index, n. A raised 1 indicates the base code is PKDGRAV while a 
raised 2 indicates GADGET. EAGLE also includes hydrodynamics to account for baryonic effects.  Equations for distribution functions or differential energy 
distributions and density profiles are given when appropriate.}
\end{table}

\begin{figure}
\centering
\includegraphics[width=\columnwidth/2]{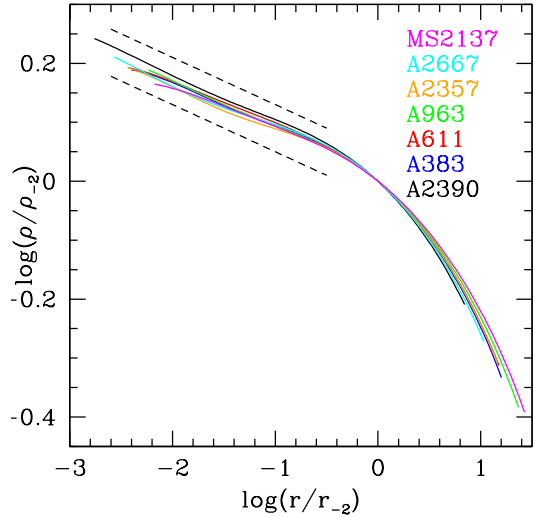}
\vskip-.6in
\caption{Density profiles for seven massive, equilibrium galaxy clusters presented in \citet{New13}. The density oscillations are the deviations from a monotonically changing 
slope in the region $\log(r/r_{-2})<0$.  The x-axis is normalized by the radius value where the density slope is $\gamma=2$ and is called $r_{-2}$.  The 
y-axis is normalized by the value of the density at $\gamma=2$, and is called $\rho_{-2}$. The dashed parallel lines serve as guidelines to better find the 
oscillations and highlight the region of interest.}
\label{Abellrho}
\end{figure}

In this paper, we concentrate on an aspect absent from NFW and other phenomenological models.  A close inspection shows that at small radii, both observations 
and high resolution simulations often exhibit oscillatory behavior in radial, total mass density profiles.  Though we call them oscillations, no time 
evolution is implied, these are time-stationary features.  Within simulations, this can be seen in the dark 
matter galactic evolution simulations GHALO and Via Lactea 2, as well as dark matter plus hydrodynamics simulations \citep{Rem13, Die08}.  
Nuclear star cluster also represent oscillations in density within the host galaxy \citep{Mer09,Bok02}. 
Observational data presented by \citet{New13} of galaxy clusters, Sloan Digital Sky Survey (SDSS) elliptical galaxies presented by \citet{Cha14}, and 
density profiles of luminous red galaxies based on satellite kinematics \citep{Tal13} show oscillations in their central regions.  
For example, Figure~\ref{Abellrho} shows the density profiles of seven equilibrium galaxy clusters presented in \citet{New13}. 
The 
density slope interior to $\log(r/r_{-2})\sim 0$ is changing non-monotonically, i.e. oscillating.  The x-axis is normalized by the largest radius value where the 
density slope is $\gamma=2$, where $\gamma=-d\log\rho/d\log r$, and is called $r_{-2}$.  The $r_{-2}$ convention is used, since in some systems, like DARKexp, 
the virial radius is not defined \citep{Wil10}.  The Newman et al. profiles are built from strong and weak gravitational lensing 
observations along with resolved stellar kinematics within the bright central galaxies to produce the radial density profiles that include both dark and 
baryonic matter.  Their profiles consist of two main components: the dark matter halo and the stars in the bright central galaxy.  For the dark matter halo, 
a generalized NFW (gNFW) profile is used, which gives the profile its $\gamma=3$ large radii behavior. 

The origins of oscillations in density profiles of equilibrium systems have not been investigated in depth. Density profiles that are monotonic in logarithmic density slope by construction, 
like NFW and Einasto \citep{Ein65}, are able to capture the general shape of halo density curves but do not capture the oscillation features in density at small radii.

Just like baryonic effects can be used to transform density cusps into cores \citep{Nav96, Gne02, Rea05, Mas08, Pon12, Tey13, Rea15, Nip15, Ono15} in the central regions of dwarfs, they can also be used as part of an 
explanation for the origins of oscillations in density profiles.  As baryons cool, they condense to the centers of halos which can further pull dark matter 
to centrally concentrate it \citep{Elz01}. Other baryonic effects, like feedback, can decrease the central concentrations in halos. All these effects change the dark matter and total matter 
configuration when compared to dark matter only systems. Hydrodynamic codes have been added to structure formation simulations to account for the competing 
baryonic effects on mass \citep{Gne04, Duf10, Ped09, Pon12, Bro12, Saw13, Saw15, Cus14, Vel14, DiC14, Sch15}.  The combination of these baryonic effects and 
the response of the dark matter usually leads to a net increase in the central mass density in clusters and early type galaxies, thereby producing 
oscillatory features in central regions. These oscillations are better described by their slopes and we will refer to density slope oscillations from now on.

However, this explanation in terms of baryons is not completely satisfactory. It does not explain why relaxation following, for example a merger, or central 
galaxy formation, or some other formation event within a cluster, does not fully incorporate the central cluster galaxy into the rest of the cluster.  In other words, 
why the change in the density slope which marks the transition in radius between the central galaxy and cluster remains easily identifiable and does not get 
erased by relaxation. The same statement can be made about the central regions of galaxies that exhibit an upturn in the mass density. The persistence of this 
visible transition implies that the relaxation in the central parts is incomplete.

In this paper we will not address the physical reasons behind incomplete relaxation; instead, we will provide support for the incomplete relaxation 
interpretation of the density slope oscillations, through two lines of argument.  First, we will show that the differential energy distributions of the 
oscillating density slope systems deviate only slightly, and in a simple way from those of fully relaxed systems, namely DARKexp. Second, we will show that 
similar, but smaller amplitude slope oscillations are also present in some high resolution pure dark matter simulated halos, and that these have differential energy 
distributions that differ from that of DARKexp in a similar way to the observed systems described above. Note that baryonic effects cannot be invoked to explain oscillating density features in 
these simulations, underscoring that baryons do not provide the full explanation in observed systems either. To put it differently, we argue that density 
slope oscillations seen in observed systems and pure dark matter simulations have a common origin: both are indicators that the central regions of these 
systems, though in stable equilibrium, are not fully relaxed. A fully relaxed system does not contain any information about its past history and assembly, and so 
should look similar to DARKexp, which does not have density slope oscillations in the radial range $-2<\log(r/r_{-2})<0$.

We call the observed and simulated systems discussed in the last three paragraphs dynamically evolved, because that is how they arrived at their present state, 
through dynamical evolution to reach their present, near steady-state configuration. Our analysis related to them (Class 2 systems) is presented in Section \ref{LocalMod}. 
However, we start our discussion of density slope oscillations in Section \ref{class1} by addressing a very different set of density profile models; those that were 
arrived at through a mathematically formulated physics-based argument or principle which we call Class 1 systems. Examples include isothermal spheres, 
DARKexp, and polytropes. The reasons for these exhibiting density slope oscillations is different than the dynamically evolved systems mentioned above.

These physics-based models start as derivations or arguments in terms of $f(E)$ or $N(E)$, i.e. quantities that involve the full dynamics, and not just the 
configuration space: these are never started as density profiles.  It appears that monotonic shapes in $f(E)$ and $N(E)$ generically lead to non-monotonic 
$\gamma(r)$ for these systems.  This is unlike phenomenological density slope profiles that are monotonically changing in log space by construction and contain 
parameter(s) that are optimized by fitting to the density profiles.  These do not have any density slope oscillations by design.  We recognize that there are 
exceptions to our classification. For example, polytrope of $n=5$ (Plummer sphere), whose density profile slope does not oscillate is a physics-based system. 
Exceptions notwithstanding, our goal is to show that oscillations in physics-based models are not an unusual occurrence.

A non-exhaustive list of different models and data with oscillating and non-oscillating density profile slopes is presented in Table 1. For 
simulations, the basic source code is identified and the softening parameter is listed. Note that for the simulation profiles with oscillations listed, the 
oscillations happen at radii larger than the softening parameter. Equations for the differential energy distribution or the distribution functions of Class 1 systems and for density profiles in the 
non-oscillating list are given when appropriate. The density profiles of non-oscillating systems in Table 1 do not have simple analytic expressions for their 
f(E) and N(E), with the exception of polytropes.  Likewise, Class 1 systems usually don't have analytic expressions for their density profiles.

While these profiles are spherically symmetric, it is important to note that actual systems could look different in three dimensions, they could have some 
ellipticity and substructure. Ellipticity means that there are density anisotropies along certain radial directions. Radial 
perturbations do not map to oscillations because the latter are tangential perturbations in the density profile. We discuss substructure at the beginning of 
Section~\ref{LocalMod}.

To summarize, not all density slope oscillations are the same; we identify two classes of systems that possess density slope oscillations: Class 1 are the 
ones seen in physical argument based models that appear to exist because most $f(E)$ and $N(E)$ lead to density profiles with non-monotonic slopes, while 
Class 2 systems are the ones seen in dynamically evolved systems that are likely to arise from incomplete relaxation.  The two causes of oscillations seem 
to be unrelated, and so can be ``superimposed'' in a single system.  Furthermore, as we discuss in the following
Sections, the two classes of oscillations affect different radial ranges in the density profiles, so they can be easily distinguished.

In Section \ref{DARKexpsec}, we describe the DARKexp distribution and its density profiles. In Section \ref{class1}, we explore additional energy distributions that, while not being 
based on physical arguments, are defined through $f(E)$ or $N(E)$ and characterize the oscillations they produce.  We then use DARKexp as a basis for fitting 
a model to observed and simulated density profiles in Section \ref{LocalMod}.  Because Class 1 and 2 systems with density slope oscillations are unrelated, Sections 3 and 
4 are largely independent of each other.  We anticipate that Class 2 systems (Section \ref{LocalMod}) will be of more interest to most readers.  We discuss our findings 
and conclude in Section \ref{Conclusions}.

\section{DARK\lowercase{exp}}\label{DARKexpsec}

DARKexp is a physics-based, equilibrium statistical mechanical theory of collisionless self-gravitating dark matter systems \citep{Hjo10}. Its 
differential energy distribution is given by
\begin{equation}\label{eq:1}
 N(\epsilon) \propto \exp(\phi_0-\epsilon)-1 
\end{equation}
where $\phi_0$ is the dimensionless central potential and $\epsilon$ is the dimensionless energy.  $\phi_0$ and $\epsilon$ can be related to $\Phi_0$ and 
$E$, the potential energy and total energy respectively, as $\phi_0 = \beta \Phi_0$ and $\epsilon=\beta E$ by $\beta$, the inverse temperature. 

The development of DARKexp is motivated by determining the origin of collisionless, self-gravitating structure. It follows similar statistical mechanics 
arguments as the derivation of Maxwell-Boltzmann distribution but deviates in two important ways.  The first is that the system is described in total orbital energy space 
instead of the traditional position and velocity 6-D phase space.  This is because particles in a collisionless system retain their energy in contrast to 
those in collisional systems.  The second point is that low occupancy numbers of the energy state space are treated more accurately with a better 
representation of $n!$ than the Stirling approximation.  

\begin{figure*}
\begin{minipage}[t]{0.45\linewidth}
\centering
\includegraphics[width=\columnwidth]{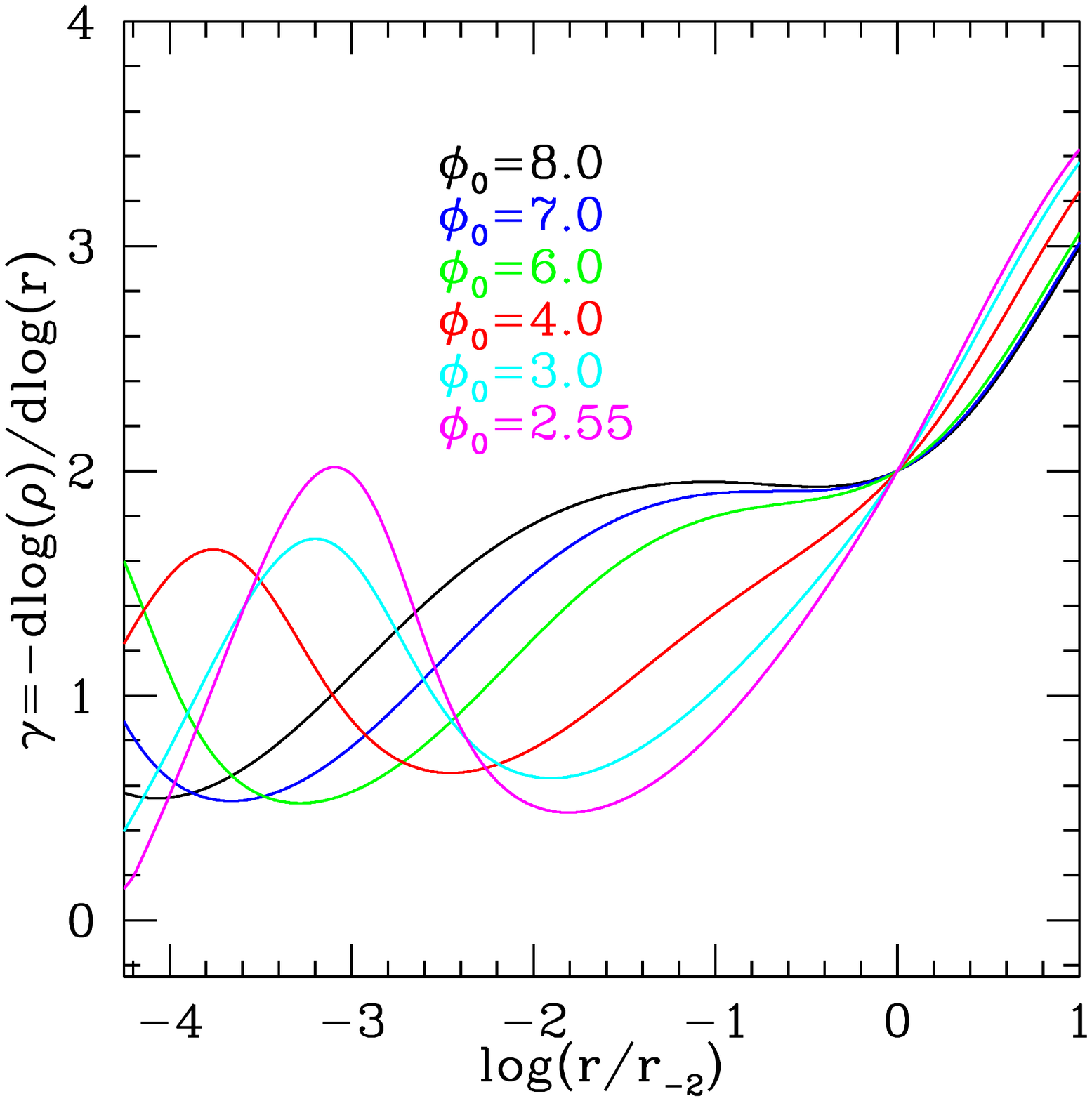}
\vskip-.6in
\caption{DARKexp density profile slopes for several central potentials. From top to bottom at $\log(r/r_{-2})=-1$, the halo potential depths are 
$\phi_0 = 8.0, 7.0, 6.0, 4.0, 3.0, 2.55$.  The horizontal axis is normalized to the radius, $r_{-2}$, where the density slope is $\gamma=2$.}
\label{DARKexp}
\end{minipage}
\hspace{0.5cm}
\begin{minipage}[t]{0.45\linewidth}
\centering
\includegraphics[width=\columnwidth]{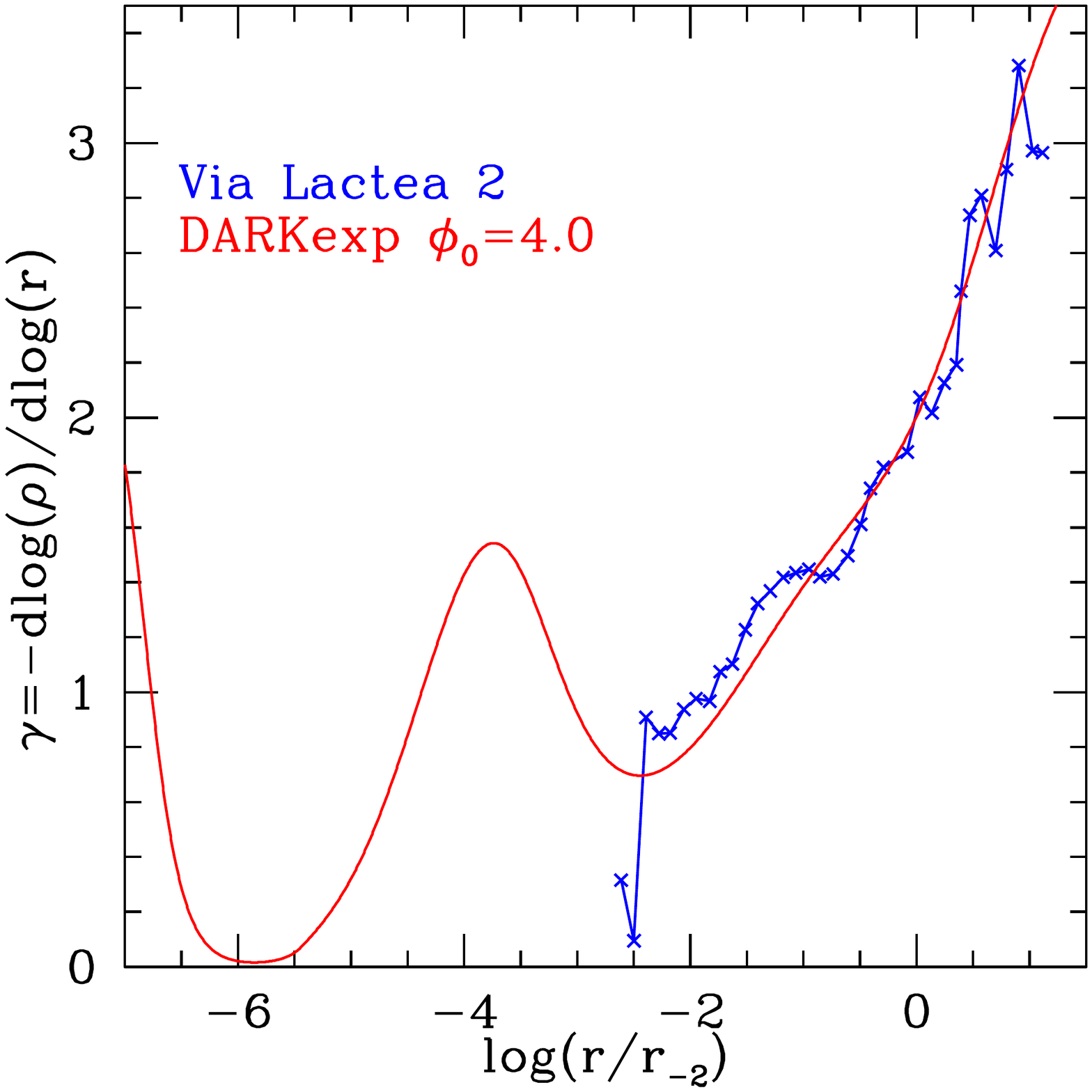}
\vskip-.6in
\caption{Density slope profile for DARKexp with $\phi_0=4.0$ and data from the simulation Via Lactea 2. Note the different amplitudes and radial regions where the 
oscillations are present.  The Via Lactea 2 oscillations are in the radial range $-2 < \log r/r_{-2} < 0$.}
\label{fullde}
\end{minipage}
\end{figure*}

While the energy distribution of DARKexp is analytically known, the resulting density profile must be calculated numerically through iteration 
\citep{Wil10,Bin82}. The DARKexp asymptotic density slope is $-1$ ($\gamma=1$) as $r\rightarrow0$ for all central potentials, but the slopes do not 
monotonically approach $-1$.  See Figure \ref{DARKexp}, where we plot the density slope, $\gamma$.  Depending on the central potential 
value, $\phi_0$, DARKexp can recreate the density profile slope range ($0\simlt\gamma\simlt 2$) seen in the above mentioned simulations and observations, at 
radii $-2\simlt \log(r/r_{-2})\simlt 0$. At these radii, DARKexp profiles that closely resemble physical systems (those with $2\simlt\phi_0\simlt 8$) do not 
exhibit slope oscillations while observed and simulated systems do; for example, see Figure~\ref{Abellrho}.

Because the oscillations in DARKexp versus those 
in dynamically evolved systems occur at different radii, there is little chance of confusing them. This can be seen in Figure~\ref{fullde} where 
density slope profile of DARKexp with $\phi_0=4.0$ and data from the simulation Via Lactea 2 are plotted together.  The Via Lactea 2 oscillations are 
in the radial range $-2 < \log r/r_{-2} < 0$.  At radii smaller than 
$\log(r/r_{-2})\sim -2$, DARKexp profiles begin to oscillate. However, these small radii are currently inaccessible to either observations or high 
resolution simulations.  We will return to DARKexp oscillations in Section~\ref{class1}.

Presumably, the oscillations extend into smaller radii but the numerical resolution of the code that produced Figure \ref{DARKexp} becomes the limiting 
factor and numerical effects start to dominate on average shortly after $\log(r/r_{-2})\sim-4.5$, although this appears to be $\phi_0$ and code resolution 
dependent. 

Aside from center oscillating $\gamma(r)$, DARKexp closely resemble results of numerical simulations and observations \citep{Sil13, Ume15, Wil14, Hjo15, Nol16} 
which do not extend much beyond $\log(r/r_{-2})\sim -2.5$.

\section{Class 1: Density Slope Oscillations in Systems Defined by $\lowercase{f}(E)$ or $N(E)$}\label{class1}

In this Section, we expand on our assertion that systems obtained from some physical argument, instead of being dynamically evolved, 
usually exhibit density slope oscillations. We also investigate if there is a link between the amplitude of oscillations in density profile slope and the 
shape of $N(E)$.  

The key aspect of physics-based systems is that their mathematical description is always for $f(E)$ or $N(E)$, and never for the density 
profile. Examples are (see Table 1): isothermal spheres, whose $f(E)$ is an exponential in energy, and whose equation of hydrostatic equilibrium 
is the same as that of isothermal gas; polytropes, whose $f(E)$ is a power law in energy, and they represent self-gravitating spheres with a polytropic 
equation of state, and DARKexp, whose derivation is described in Section \ref{DARKexpsec}. 

All above examples, as well as other examples of physics-based models have simple analytical expressions for either $f(E)$ or $N(E)$. This is their distinguishing feature. On the other hand, non physics-based models have either very complicated expressions for $f(E)$ or $N(E)$ because they were obtained from $\rho(r)$ (like Hernquist, 1990), or do not have analytical expressions at all (like NFW and Einasto).

While all the Class 1 systems in Table 1 oscillate, not all do so at small radii. Polytropes at $n$ around 5 have density slope oscillations at large radii \citep{Med01}, while isothermal spheres have them at intermediate radii \citep{Bin87}. 
Because the model that represents relaxed collisionless systems, DARKexp, has oscillations at very small radii, we are mostly interested in small radius oscillations, and therefore we decided to use simple functional forms of $N(E)$ that produce density profiles which oscillate at small radii.

We chose two functional forms:
\begin{equation}\label{eq:2}
N(\epsilon)\propto (\phi_0-\epsilon)^{1/\alpha}
\end{equation}
and
\begin{equation}\label{eq:3}
 N(\epsilon)\propto e^{\phi_0-\epsilon}(\phi_0-\epsilon)^{1/\alpha}
\end{equation}
where $\phi_0$ is a dimensionless central potential found in DARKexp and $\alpha$ is a free parameter.

Though these $N(E)$ are not a result of a physical principle or argument, they share one key aspect with isothermal spheres, polytropes, and DARKexp: they are defined through simple expression of $f(E)$ or $N(E)$.  We now show that for these systems, density slope oscillations are common and that the shape of $N(E)$ can approximately predict the 
``strength" of oscillations of the corresponding density profiles.

\begin{figure}
\centering
\includegraphics[width=\columnwidth/2]{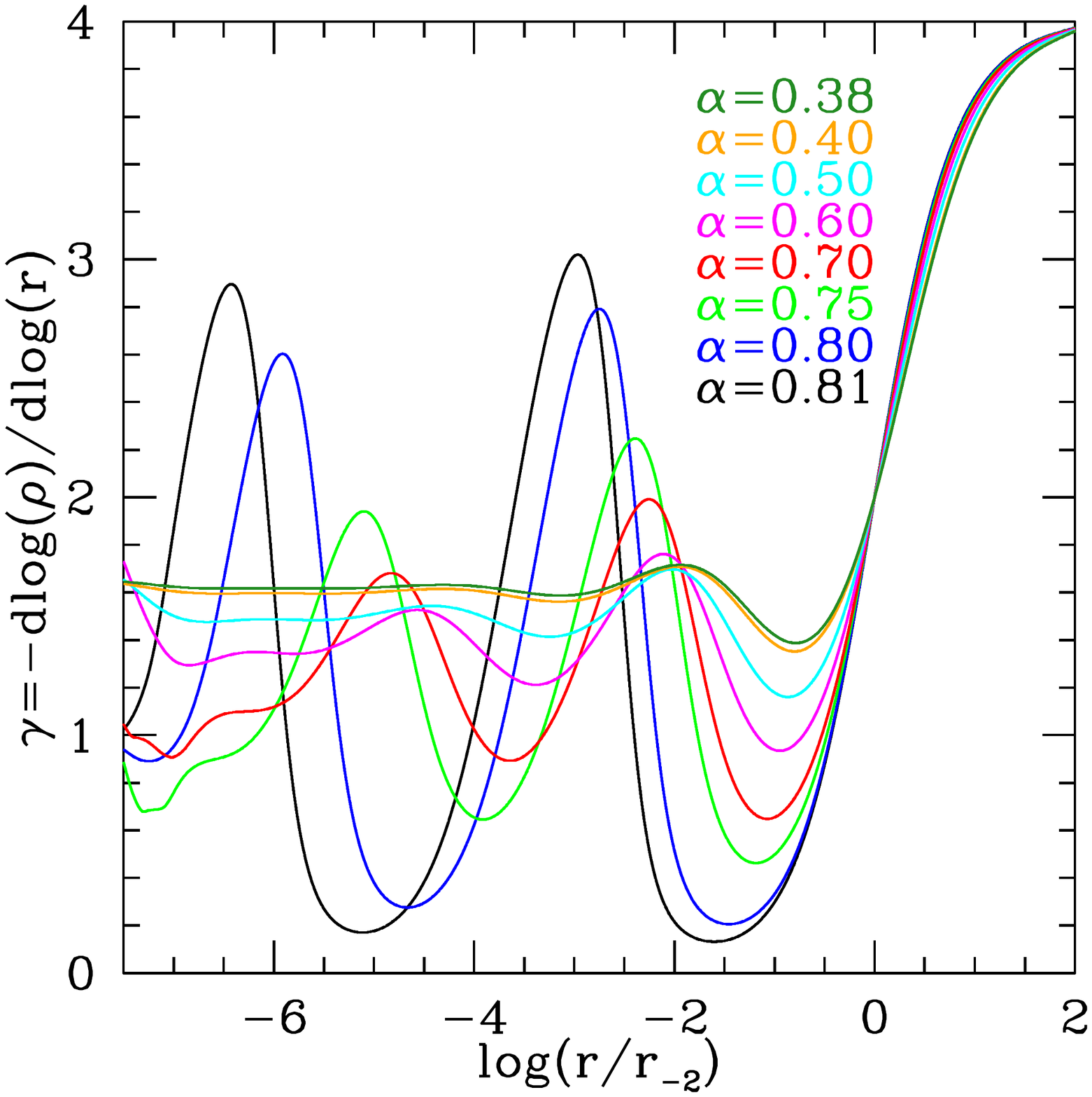}
\vskip-.6in
\caption{Density profile slopes created from energy distributions given by $N=(\phi_0-\epsilon)^{1/\alpha}$ with $\phi_0=4.0$. The values of $\alpha$ 
span a range that provides numerical convergence in the code. The higher the value of $\alpha$, the larger the oscillation in density profile. From top to 
bottom at $\log(r/r_{-2})=-1$, $\alpha= 0.38, 0.40, 0.50, 0.60, 0.70, 0.75, 0.80, 0.81$. The asymptotic log-log density profile slope at small radii is 
given by $\gamma \propto 2-\alpha$, which yields $\gamma_{asymp}= 1.62, 1.6, 1.5, 1.4, 1.3, 1.25, 1.2, 1.19$ for the previous list of $\alpha$ values, respectively.  
These values will define a baseline used for further comparisons.}
\label{acurve1_40}
\end{figure}

\begin{figure}
\centering
\includegraphics[width=\columnwidth/2]{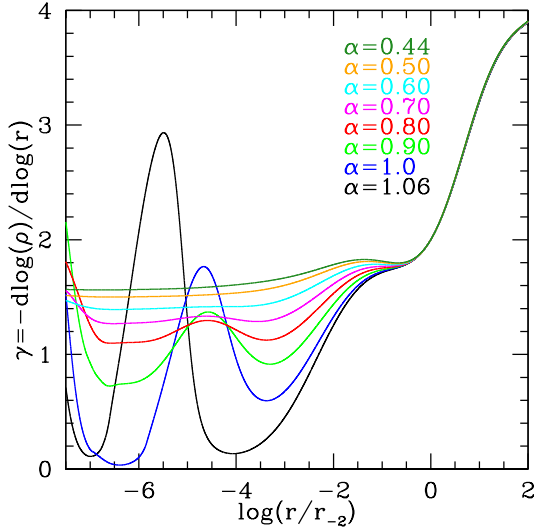}
\vskip-.6in
\caption{Density profile slopes created from energy distributions given by $N=e^{\phi_0-\epsilon}(\phi_0-\epsilon)^{1/\alpha}$ with $\phi_0=4.0$. The 
values of $\alpha$ span a range that provides numerical convergence in the code.  The higher the value of $\alpha$, the larger the oscillation in density 
profile. From top to bottom at $\log(r/r_{-2})=-2$, $\alpha= 0.44, 0.50, 0.60, 0.70, 0.80, 0.90, 1.0, 1.06$. The asymptotic log-log density profile slope at small radii is 
given by $\gamma \propto 2-\alpha$, which yields $\gamma= 1.56, 1.5, 1.4, 1.3, 1.2, 1.1, 1.0, 0.94$ for the previous list $\alpha$ values, respectively.  
These values will define a baseline used for further comparisons.}
\label{acurve2_40}
\end{figure}

Two sample profile collections are shown in Figure \ref{acurve1_40} and Figure \ref{acurve2_40} and represent equation \eqref{eq:2} and equation \eqref{eq:3},
respectively; both assume a central potential of $\phi_0=4.0$.  The range of $\alpha$ values ensured convergence given the code's resolution in radius. For 
all the $\alpha$ values for each energy distribution, the density seems to oscillate about the expected small radii behavior given in \citet{Hjo10}, 
which is $\rho \propto r^{\alpha -2}$, but with different amplitudes.

\begin{figure}
\centering
\includegraphics[width=\columnwidth/2]{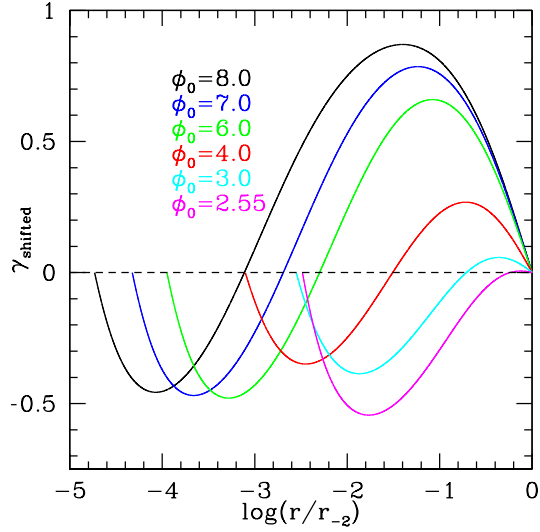}
\vskip-.6in
\caption{One oscillation of the density profile slopes from Figure \ref{DARKexp} after their small radius asymptotic 
baseline has been subtracted which for DARKexp is $\rho \propto r^{-1}$. The positive area was calculated from these profiles. The term ``positive area'' means the absolute value was 
applied for the parts of the curve that would produce negative area. We called this area under one oscillation, $\kappa$. For reference, the black dashed 
line is $\gamma=0$.}
\label{DARKexponeosc}
\end{figure}

To quantitatively assess the density profiles, we created an oscillation parameter, $\kappa$. We will use $\kappa$ to quantify oscillations
as it gauges the difference between the profile slope and its asymptotic small radius behavior.  Since $\rho \propto r^{\alpha -2}$, this would 
mean $\gamma \propto 2-\alpha$ at small radii in log space.  $\kappa$ is defined as the enclosed area between the density profile slope, $\gamma$, and 
the requisite baseline $2-\alpha$.  $\kappa$ was calculated over one complete oscillation for each 
$\alpha$ value. One oscillation is counted as crossing the small radius baseline two additional times starting from $\log(r/r_{-2})=0$ and can be seen in 
Figure \ref{DARKexponeosc} for DARKexp of several $\phi_0$.  

There are some limitations to using this area characterization.  It may not accurately represent high frequency oscillations that could be defined as 
oscillating more than a low frequency oscillation with the same amplitude.  Fortunately, we did not see any oscillations that had different frequencies but the same amplitude.  
The concept of frequency used here is the measure of how often the peak in the oscillation is repeated per unit distance in $\log(r/r_{-2})$ space.

The curves in Figure \ref{DARKexponeosc} represent the result of the subtraction between the original $\gamma$ profile and the asymptotic behavior for DARKexp 
profiles, which is $\rho \propto r^{-1}$, so $\gamma=1$.  The result is then integrated to find the area, $\kappa$; all area is considered positive.  

\begin{figure}
\centering
\includegraphics[width=\columnwidth/2]{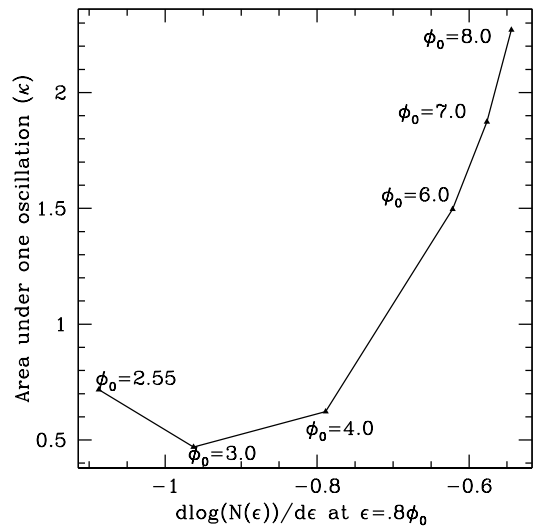}
\vskip-.6in
\caption{Area under the curve, $\kappa$, for one oscillation of DARKexp central potentials $\phi_0=8.0, 7.0, 6.0, 4.0, 3.0, 2.55$. The slope of the density 
profile had to be evaluated at a specific energy, so $\epsilon=0.8\phi_0$ was arbitrarily chosen.}
\label{DARKexparea}
\end{figure}

Our aim is to relate the amplitude of density slope oscillations to the shape of the differential energy distribution. From visual examination, we noticed 
that $N(E)$ which are shallower (have smaller absolute values of $d\log(N)/d\epsilon$) close to the most binding energy tend to produce density profiles with 
higher slope oscillations. Trial and error showed that taking the slope of $N(E)$ at $\epsilon=0.8\phi_0$ works well. Figure \ref{DARKexparea} plots the area under the 
curve, $\kappa$, against $d\log(N)/d\epsilon$ evaluated at $\epsilon=0.8\phi_0$. The plotted curve shows that there is a relation between slope oscillations 
in $\rho(r)$ and $N(E)$, or that the shape of the energy distribution determines how much the density slope will oscillate.

\begin{figure*}
\begin{minipage}[t]{0.45\linewidth}
\centering
\includegraphics[width=\columnwidth]{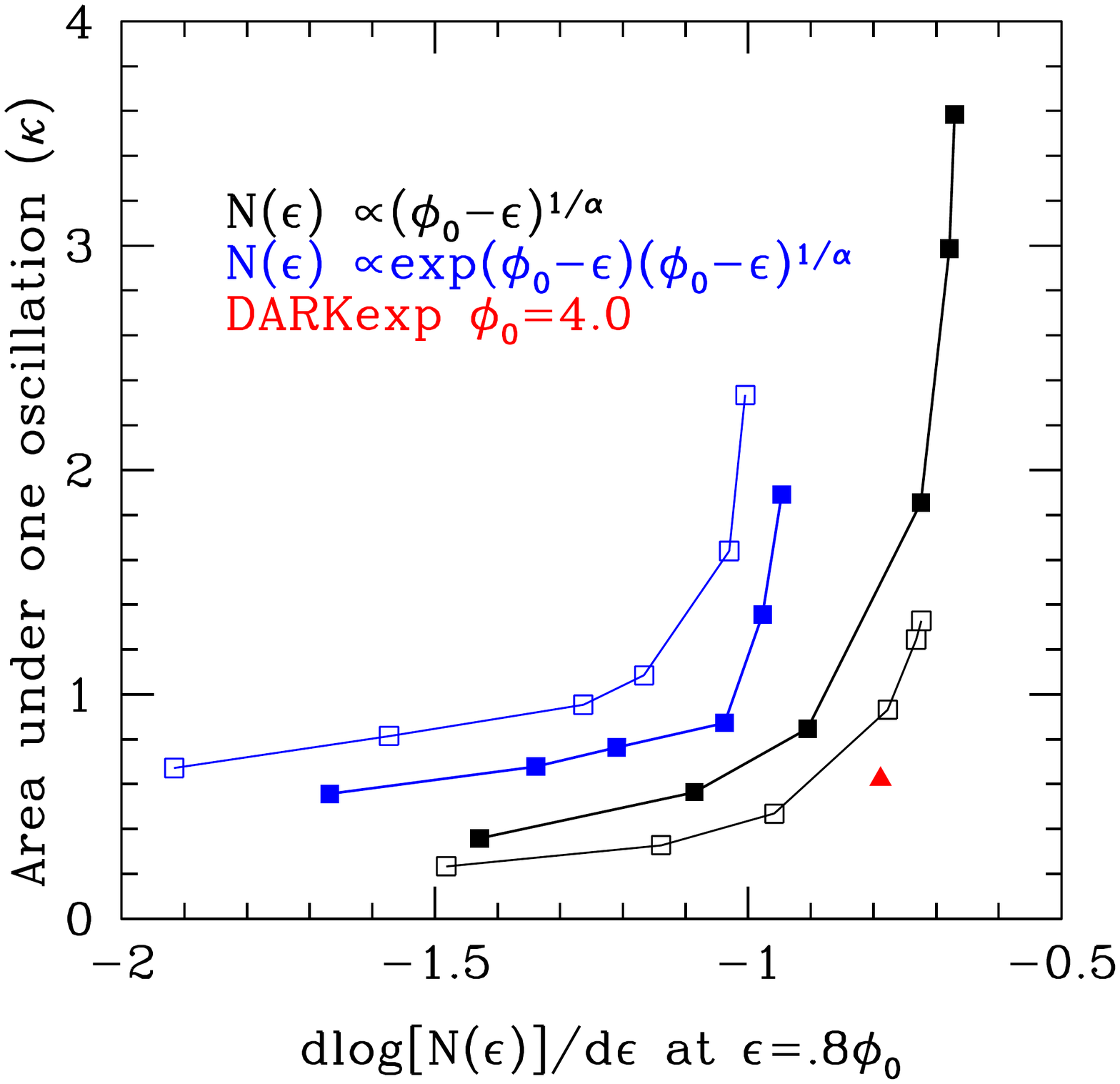}
\vskip-.6in
\caption{Area under the curve, $\kappa$, for one oscillation for $N=e^{\phi_0-\epsilon}(\phi_0-\epsilon)^{1/\alpha}$ and $N=(\phi_0-\epsilon)^{1/\alpha}$ with 
$\phi_0=4.0$. The different points represent the different $\alpha$ values with (from top to bottom) $\alpha=1.05, 1.0, 0.8, 0.7, 0.5, 0.38$ for 
$N=e^{\phi_0-\epsilon}(\phi_0-\epsilon)^{1/\alpha}$ and with (from top to bottom) $\alpha=0.81, 0.8, 0.75, 0.6, 0.5, 0.38$ for $N=(\phi_0-\epsilon)^{1/\alpha}$. 
Solid squares represent the original $N=e^{\phi_0-\epsilon}(\phi_0-\epsilon)^{1/\alpha}$ and $N=(\phi_0-\epsilon)^{1/\alpha}$ distributions and open squares 
represent the distributions modified to be $N_{new} = N(1+\mu(\phi_0-\epsilon)^v)$.  DARKexp with $\phi_0=4.0$ is plotted as the red data point for reference.}
\label{area40}
\end{minipage}
\hspace{0.5cm}
\begin{minipage}[t]{0.45\linewidth}
\centering
\includegraphics[width=\columnwidth]{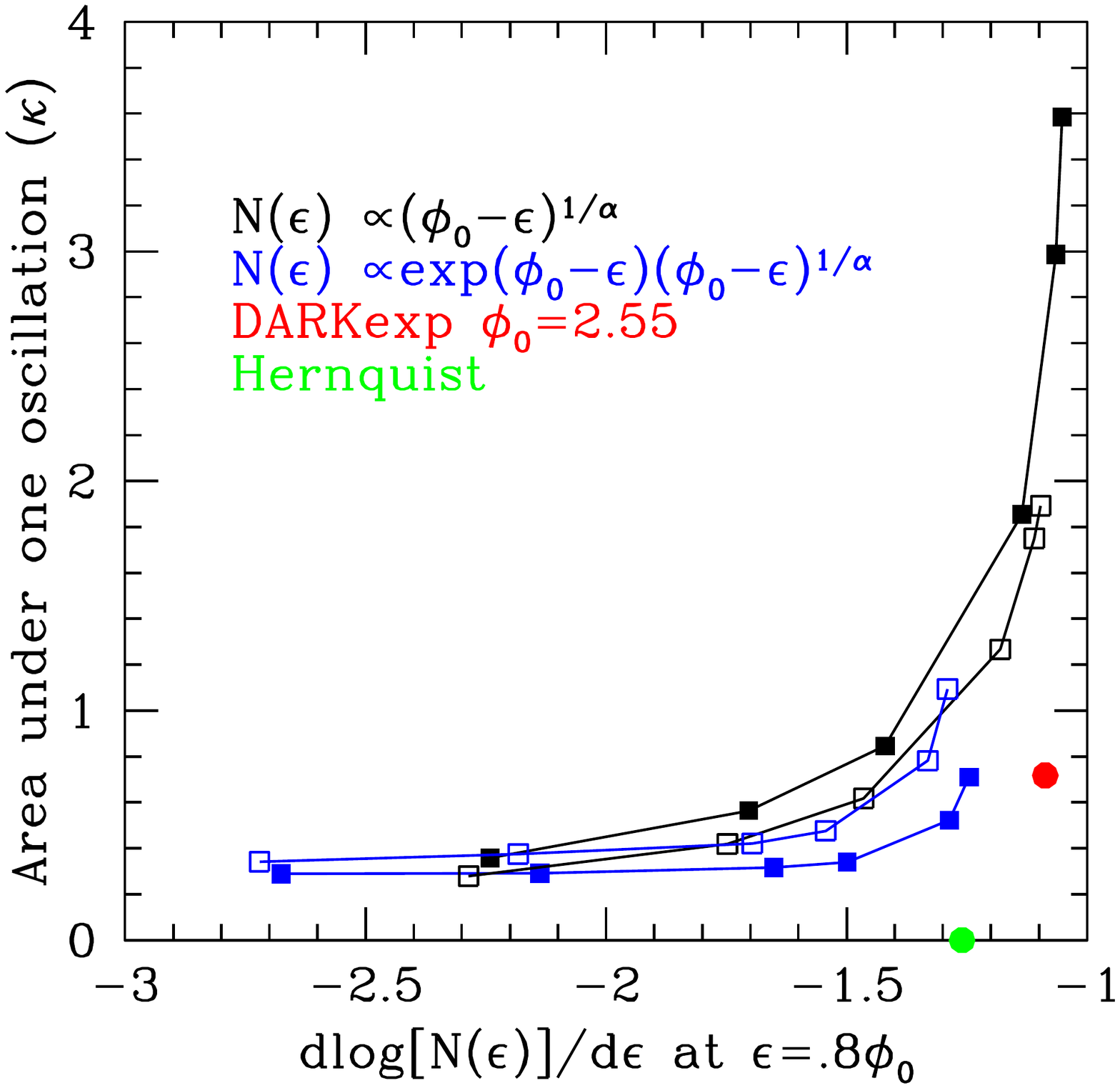}
\vskip-.6in
\caption{Area under the curve, $\kappa$, for one oscillation for $N=e^{\phi_0-\epsilon}(\phi_0-\epsilon)^{1/\alpha}$ and $N=(\phi_0-\epsilon)^{1/\alpha}$ with 
$\phi_0=2.55$. The different points represent the different $\alpha$ values with (from top to bottom) $\alpha=1.05, 1.0, 0.8, 0.7, 0.5, 0.38$ for 
$N=e^{\phi_0-\epsilon}(\phi_0-\epsilon)^{1/\alpha}$ and with (from top to bottom) $\alpha=0.81, 0.8, 0.75, 0.6, 0.5, 0.38$ for $N=(\phi_0-\epsilon)^{1/\alpha}$. 
Solid squares represent the original $N=e^{\phi_0-\epsilon}(\phi_0-\epsilon)^{1/\alpha}$ and $N=(\phi_0-\epsilon)^{1/\alpha}$ distributions and open squares 
represent the distributions modified to be $N_{new} = N(1+\mu(\phi_0-\epsilon)^v)$.  DARKexp with $\phi_0=2.55$ is plotted as the red data point for reference.}
\label{area255}
\end{minipage}
\end{figure*}

We did the same analysis for the different functional families given by equation \eqref{eq:2} and equation \eqref{eq:3} with $\alpha=1.05, 1.0, 0.8, 0.7, 0.5, 0.38$ 
and $\alpha=0.81, 0.8, 0.75, 0.6, 0.5, 0.38$ respectively.  $\kappa$ for both of the two families with $\phi_0=4.0$ $(\phi_0=2.55)$ are plotted as solid boxes in
Figure \ref{area40} (Figure \ref{area255}) as well as a red data point for DARKexp with the same $\phi_0$.  Across both families and both central 
potentials, $\kappa$ increases as $\alpha$ increases. It appears that less negative slopes, $d\log(N)/d\epsilon$ (higher $\alpha$), in energy 
density produce more area under the curve and therefore more oscillations in density profiles for these two distributions, by our metric. 

From the fact that $\kappa$ increases with increasing $\alpha$, one might think that the oscillation amplitude would be tied to asymptotic slope but this is 
only true for the energy distributions in equation \eqref{eq:2} and equation \eqref{eq:3} and not true for DARKexp.

To further test the idea of the oscillation's dependence on the steepness of the energy distribution slope, we altered $N=e^{\phi_0-\epsilon}(\phi_0-\epsilon)^{1/\alpha}$ 
and $N=(\phi_0-\epsilon)^{1/\alpha}$ by multiplying these energy distributions so that $N_{new} = N(1+\mu(\phi_0-\epsilon)^v)$ where $N$ is the original 
distribution and $\mu$ and $v$ are free parameters. The shape of the modification is arbitrary; we could have chosen a different modification.  This resulted 
in the energy distribution having the same behavior near $\epsilon=\phi_0$ as the original distribution but a slightly steeper or shallower slope depending 
on the sign of $\mu$ and the normalization of the new distribution. We then calculated $\kappa$ for both of the modifications and plotted them as open boxes 
alongside their counterparts in Figure \ref{area40} and Figure \ref{area255} for modifications with central potentials of $\phi_0=4.0$ and $\phi_0=2.55$, 
respectively.

Figures~\ref{DARKexparea}, \ref{area40}, and \ref{area255} show that in general less negative $d\log(N)/d\epsilon$ lead to more pronounced oscillation in the density profile slope. However, $d\log(N)/d\epsilon$ is not an exact predictor of density slope oscillations: in Figures \ref{area40} and \ref{area255} a given value of $d\log(N)/d\epsilon$ corresponds to a range of values of $\kappa$. With more effort, it might be possible to devise a better predictor of density slope oscillations based on $N(E)$, but that is beyond the exploratory nature of the present paper. Furthermore, there are exceptions to our characterization shown in Figures \ref{area40} and \ref{area255}: the shape of $N(\epsilon)$ of the Hernquist system \citep{Her90} would predict that its density profile slope should show some oscillations, which is not the case. The corresponding point in Figure~\ref{area255} is at approximately $(-1.25,0)$.

\section{Class 2: Density Slope Oscillations in Dynamically Evolved Halos} \label{LocalMod}

Here we show that density profiles of simulated and observed systems are well
approximated by those obtained from modified DARKexp energy distributions. The
modifications to DARKexp are small, implying that these simulated and observed
systems are very close to being fully relaxed.

Our analysis implicitly assumes that the simulated and observed systems we consider
here are the result of mostly collisionless relaxation. Here, the term collisionless
applies to the gravitational aspect of the system's evolution, not to what its gas
particles might be doing. To be gravitationally collisionless the potential must be
smooth on the scale of the relevant particles. In the case of the centers of
galaxies and clusters these particles are stars, and to make the potential grainy,
and hence collisional, the number of stars must be considerably smaller than what
these galaxies have. Hydrodynamic processes that operate in massive galaxies and
clusters, including dissipation, may lead to ``violent'' potential fluctuations that
can help drive collisionless relaxation. Therefore the evolution of both simulated
pure dark matter halos, and massive observed systems can be considered
collisionless.

\subsection{Numerically simulated systems}

\begin{figure*}
\centering
\includegraphics[scale=.68 ]{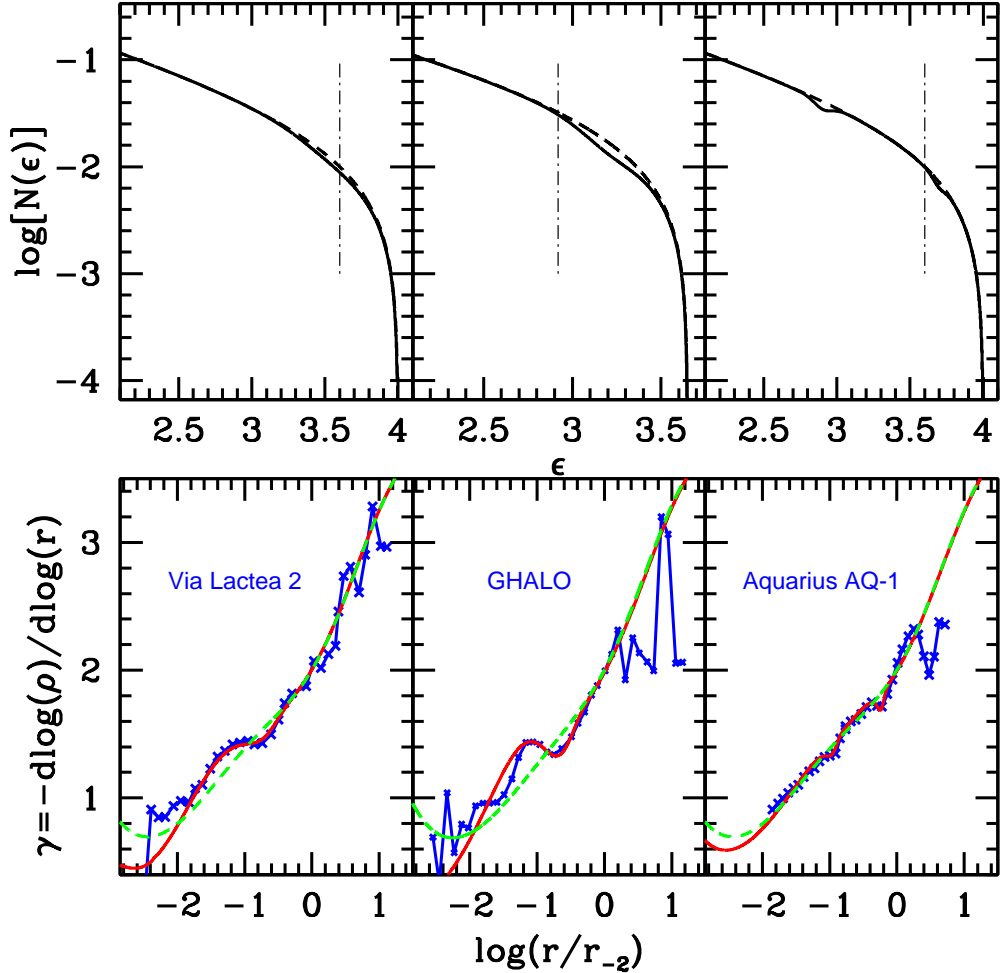}
\vskip-1.5in
\caption{{\it Top panels} Energy distributions for Via Lactea 2 fit (left), GHALO fit (middle), and Aquarius AQ-1 fit (right) in solid lines plotted with DARKexp, dashed line, with the appropriate $\phi_0$ where for Via Lactea $\phi_0=4.0$, GHALO $\phi_0=3.65$ , and Aquarius AQ-1 $\phi_0=4.0$. The slight deviation is distinguishable at around $\epsilon=0.8\phi_0$ where the two lines separate and a dashed-dot guideline is plotted at $\epsilon=0.8\phi_0$ for reference. The difference between the two curves is a Gaussian peak and in the case of Aquarius AQ-1, two Gaussian peaks.  The graph is zoomed in to this region to better show the difference between the curves and does not include the entire energy distribution. {\it Bottom panels} Density profile slopes for Via Lactea 2, GHALO, and Aquarius AQ-1 numerical simulations are plotted in blue and x symbols.  The modified DARKexp that fits the data is plotted in red and the original DARKexp is plotted in green (dashed). The modified DARKexp distributions use the $\phi_0$s listed above. In other words, the red line is a modified form of the green line. The bottom plots correspond to the plots directly above.  The mass difference in the profile for $\log(r/r_{-2})<0$ between DARKexp and the modification is 0.58\%, 0.56\%, and 0.77\% for Via Lactea 2, GHALO, and Aquarius AQ-1 respectively.}
\label{6panel}
\end{figure*}

The density profile of DARKexp with $\phi_0\approx4.0$ provides a good fit to the density profiles of the dark matter only simulations Via Lactea 2, GHALO \citep{Sta09}, and Aquarius AQ-1 \citep{Spr08} as shown in the bottom panels of Figure \ref{6panel}. These simulations are high resolution and focus on halo central regions with the GHALO and Via Lactea 2 innermost resolved region being 0.05\% of $R_{vir}$ (120 pc) \citep{Sta09}. However, DARKexp is not a perfect match; there appear to be oscillations with higher frequencies in GHALO and Via Lactea 2 data, and at larger radii than the oscillations experienced in DARKexp density profiles.  Aquarius AQ-1 does not appear to oscillate appreciably and is actually very well represented by DARKexp. 

It is important to note that these oscillations are not the result of substructure. Substructure effects are subdominant to those of oscillations (if oscillations are present), as can be seen from published work. First, data from \citet{Sta09}, can be used to estimate the fractional density contained in oscillations in GHALO; it is about 10\%. Oscillations are found around $r\sim 0.3-20$ kpc in their galaxy mass halo. Second, at the same range of radii, the cumulative fraction of mass in subhalos for a similar mass halo in the Aquarius simulation can be read off from Figure 12 of \citet{Spr08}. It is about 0.1\%, or 100 times smaller than the mass contained in oscillations found in GHALO. Aquarius and GHALO are different simulations, but it is unlikely that they would differ by a factor of 100 in the fractional amount of mass in subhalos.  Another possible reason for the oscillations are the streams from tidally disrupted subhalos. However, because the material from these is widely distributed throughout the halo, they are unlikely to be the dominant contributor to the oscillations.

We argue that oscillations in Via Lactea 2 and GHALO indicate that these systems are not fully relaxed. DARKexp assumes a relaxed, equilibrium system. This central assumption may not be fully realized in many physical systems where some mechanism may not be letting the system become fully relaxed. To try to mimic this effect, we modified DARKexp $N(\epsilon)$ with the goal of matching the observed density profile of these simulated halos.  We used the simplest profile modification: a small Gaussian peak was added to DARKexp.  The new $N(\epsilon)$ is given by
\begin{equation}
N_{mod} = N+a\exp\left[-\dfrac{1}{2}\dfrac{(\epsilon-\bar{\epsilon})^2}{\sigma^2}\right]
\label{Nmod}
\end{equation}
where $N$ is the standard DARKexp expression given by equation~\eqref{eq:1}, $a$ is a normalization scaling factor, and $\bar{\epsilon}$ and $\sigma$ are the standard Gaussian parameters of average value and standard deviation, respectively.  The modification can be seen in the top panel of Figure~\ref{6panel}, where $\log[N(\epsilon)]$ is plotted for the original DARKexp energy distribution (dashed line) and the energy distribution modified from DARKexp that provide good fits to the density data (solid lines).

The corresponding density profiles are seen as red solid lines in the bottom panels of Figure~\ref{6panel}, together with the data for Via Lactea 2 (left), 
GHALO (middle), and Aquarius AQ-1 (right), displayed as blue lines with crosses. Dashed green lines are the unmodified DARKexp profiles. In Via Lactea 2 and 
GHALO, this modification produced oscillations over two decades in radius with higher frequency than the oscillations found in the unmodified DARKexp 
profile, and at radii where unmodified DARKexp has a monotonically changing $\gamma(r)$ profile. The Aquarius modification, two very small Gaussian peaks 
added to DARKexp $N(\epsilon)$, did provide a better fit to the data then DARKexp alone but the Aquarius data does not oscillate like the Via Lactea 2 or 
GHALO data. The mass difference of the profiles for $\log(r/r_{-2})<0$ between the modified and unmodified DARKexp profiles with the same $\phi_0$ is 
0.58\%, 0.56\%, and 0.77\% for Via Lactea 2 ($\phi_0=4.0$), GHALO ($\phi_0=3.65$), and Aquarius ($\phi_0=4.0$) respectively. This is a small difference 
confined to small radii. Again, one could imagine that in some systems, the oscillations will get averaged out by anisotropies or mis-centering and spherical averaging.  
Likewise, it would not take much mass to create oscillations in this radius range, implying that these systems are very nearly, but not fully relaxed.

While we do not address the specific mechanisms leading to the small perturbation in the DARKexp energy distribution, we put forth one possibility. The systems which have $N(E)$ represented by the solid lines in the top panels of Figure~\ref{6panel} are prevented from achieving a fully relaxed state, represented by the dashed lines of the DARKexp $N(\epsilon)$. Because the number of particles increases rapidly with decreasing $\epsilon$, it would take a very small number of particles with $\epsilon\lesssim 3$ (for Via Lactea 2) to fill in the `hole' at  $\epsilon\sim 3.5$. These particles would need to lose some energy and become more bound. What is holding them back?  For example, small excess of angular momentum could create an angular momentum barrier preventing the particles from coming closer to the center of the halo and attaining lower energies. Such a scenario is not unlikely because, as discussed in \citet{Wil14} it is more difficult to redistribute angular momentum than energy, and so some particles may be stuck with their somewhat higher amount of angular momentum.

\subsection{Observed systems}

\begin{figure}
\centering
\includegraphics[width=\columnwidth/2]{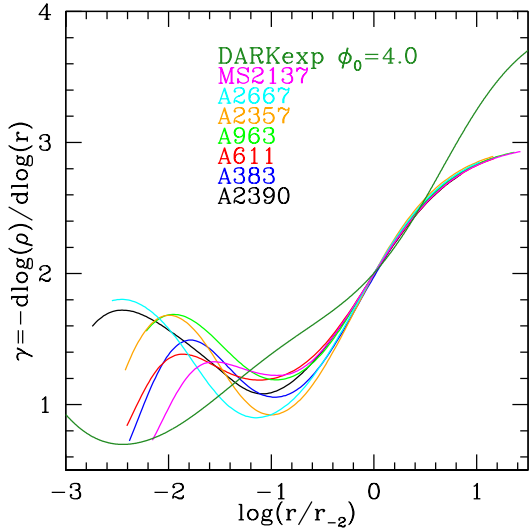}
\vskip-.6in
\caption{Density profile slopes for seven massive, equilibrium galaxy clusters presented in \citet{New13}. Oscillations are seen for all the profiles for 
$\log(r/r_{-2})<0$. DARKexp with $\phi_0=4.0$ is plotted for comparison. The Newman profiles asymptote to the outer slope of 3.  The oscillations are higher 
frequency for the Newman profiles than DARKexp.}
\label{Abell}
\end{figure}

The observationally derived data taken from \citet{New13} provides density profiles for seven massive, equilibrium galaxy clusters seen in Figure~\ref{Abellrho}. We transformed this data into the slope representation, $\gamma$, and plotted it in Figure \ref{Abell} to better see the oscillations.  All of the profiles possess oscillations in the central region. For reference, the density from a DARKexp profile with $\phi_0=4.0$ is also plotted.  At large radii, the differences between DARKexp and the data arise because \citet{New13} dark matter component is approximated with a gNFW profile which has $\gamma=3$ at large radii, while DARKexp asymptotes to $\gamma=4$.  We do not use the outer density profile in our analysis.

In addition to cluster total density profiles, we also use profiles of massive ellipticals, taken from \citet{Cha14}, who carried out a comprehensive study of SDSS early type galaxies (ETGs) that also show density slope oscillations, or as they call them, S-shape profiles, at small radii. They used over 2,000 nearly spherical ETGs from SDSS coupled with empirical scaling relations to derive galaxy density profiles. The observational data for each galaxy consisted of its light profile and the velocity dispersion measured at $R_{\rm{eff}}/8$. The dark matter halos of galaxies were assumed to follow gNFW or Einasto profiles, each with three free parameters. These data and assumptions were supplemented with empirical scaling relations between $M_{\odot}$ and $M_{200}$, concentration and $M_{200}$, initial mass function (IMF) shape and the velocity dispersion, and velocity dispersion and the slope of the radially dependent velocity dispersion profile. The free parameters were fixed by requiring that the total density profile obey the Jeans equation. The authors present average density profiles for six separate total mass bins. We note that according to the \citet{Cha14} results, the amplitude of oscillations increases towards the higher galaxy mass bins. The oscillations in the density slopes of $M=10^{14.5} M_\odot$ galaxies will be hard to fit with a small modification we use here. However, the overall nature of slope oscillations is the same for the whole range of galaxy masses presented in \citet{Cha14}, from  $M=10^{12} M_\odot$ to  $M=10^{14.5}M_\odot$, so below we use just one representative mass range, $10^{13}M_{\odot} \pm0.2 $dex.

Since we are analyzing density slope oscillations, it is important to establish that cluster and galaxy data do indeed require oscillations to be present. Could one have instead fitted the data with a monotonically changing $\gamma(r)$ profile?  \citet{Cha14} tried that approach, and found that single density profiles do not work (see their Section 4.2.1). In Figure 9, they plot the observed and single-profile predicted distributions of $\eta$, where $\sigma(R)\propto R^\eta$, is the luminosity weighted line of sight velocity dispersion within aperture $R$ centered on the galaxy. The data gives $\langle\eta\rangle=-0.06\pm0.01$, while fitting a single gNFW yields $\langle\eta\rangle=-0.01$, or $5\sigma$ away from the observed mean. \citet{New13} do not fit single profile models, but point out that the stellar mass of the central elliptical is significant at radii $\simlt 5$ kpc, often leading to the steepening of the central profile slope. We conclude that for cluster and galaxy total density profiles, single-profile models are statistically disfavored and hence, density slope oscillations are indicated by the data.

\begin{figure}
\centering
\includegraphics[width=\columnwidth/2]{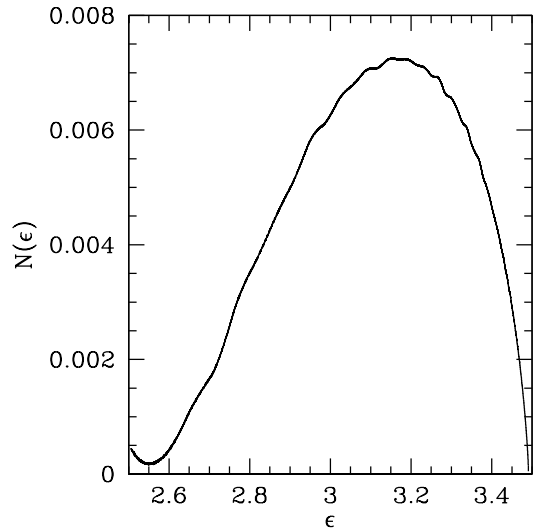}
\vskip-.6in
\caption{Difference between $N(E)$ derived from galaxy cluster Abell 611 (A611) density profile data and DARKexp $N(\epsilon)$ for $\phi_0=3.5$.  The energy distribution for A611 is calculated from the density profile with a process outlined by \citet{Bin82}. The peak is similar to a Gaussian, warranting the use of a Gaussian modification 
in this Section.}
\label{Diff peak}
\end{figure}

\begin{figure*}
\begin{minipage}[t]{0.45\linewidth}
\centering
\includegraphics[width=\columnwidth]{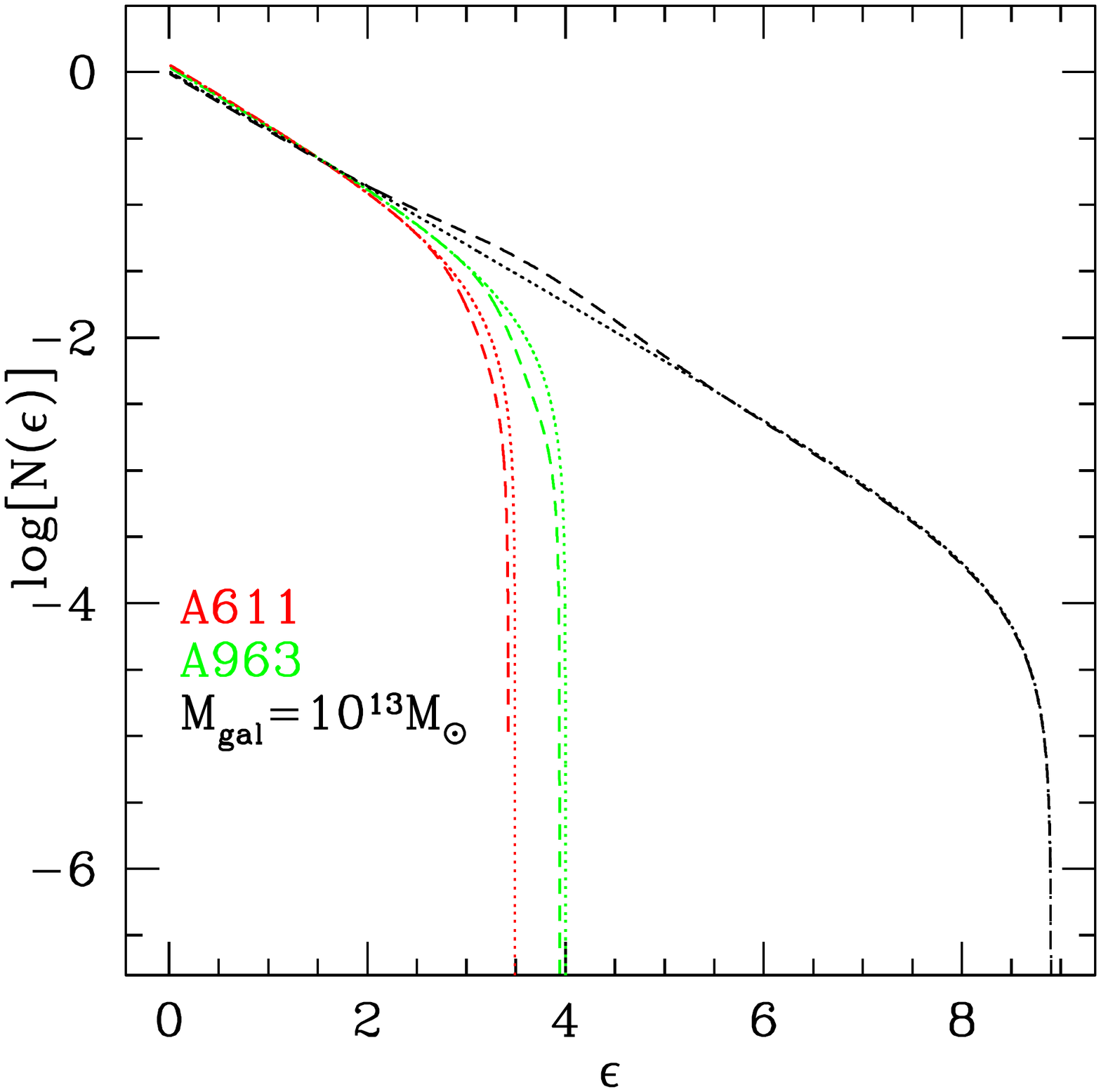}
\vskip-.6in
\caption{Energy distributions of the modified DARKexp profiles (dashed) plotted against their original DARKexp profiles (dotted) with $\phi_0=3.5$, 
$\phi_0=4.0$, and $\phi_0=8.9$. The deviations from DARKexp are distinguishable at around 85\% of $\phi_0$ for A611 and A963 whereas the peak for the 
Gaussian for the \citet{Cha14} $M_{gal}=10^{13}M_{\odot}$ modification is found at 35\% of $\phi_0$. The modified DARKexp profiles provide fits to the density 
profiles plotted in Figure \ref{Observation Fits}.
}
\label{Observation Nplots}
\end{minipage}
\hspace{0.5cm}
\begin{minipage}[t]{0.45\linewidth}
\centering
\includegraphics[width=\columnwidth]{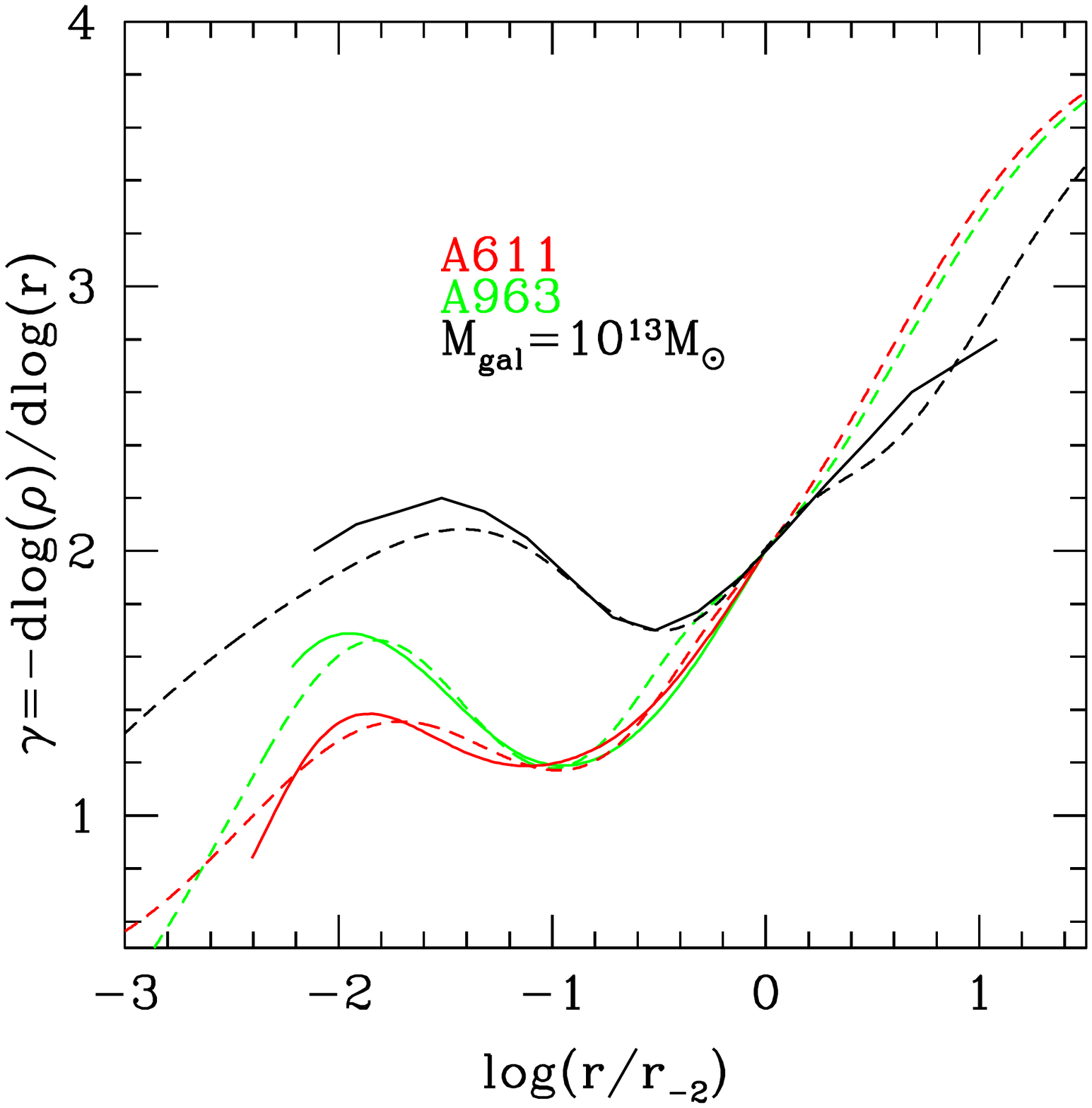}
\vskip-.6in
\caption{Density profile slopes for three profiles (solid lines), one taken from \citet{Cha14} and two from \citet{New13}, and their best fit formed from 
modifying DARKexp with a Gaussian (dashed lines). The mass difference in the profiles at $\log(r/r_{-2})<0$ between DARKexp and the modification is 2.0\%, 
2.52\%, and 19.1\% for \citet{New13} A611, A963 and \citet{Cha14} $10^{13}M_{\odot}$ respectively. The modified profiles (dashed) are the density profiles 
that correspond to the modified DARKexp energy distributions found in Figure \ref{Observation Nplots}.
}
\label{Observation Fits}
\end{minipage}
\end{figure*}

We chose two galaxy clusters from \citet{New13}, A963 and A611, and one galaxy mass range from \citet{Cha14}, $10^{13}M_{\odot} \pm0.2$dex. The unmodified DARKexp (dotted) and modified (dashed) $N(\epsilon)$ are shown in Figure~\ref{Observation Nplots} with green, red, and black lines respectively. The corresponding modified $\gamma(r)$ profiles are the dashed lines in Figure~\ref{Observation Fits}, shown together with the data (solid lines). For fitting purposes, we only used data where $\log(r/r_{-2})<0$ since DARKexp and the data have different large radius asymptotic behavior, as discussed above. 

In all the fits, the introduction of a Gaussian modification to $N(\epsilon)$ modified DARKexp density profiles such that these now fit the frequency and amplitude of the observed data. The mass difference in the profiles at $\log(r/r_{-2})<0$ between the modified and their original DARKexp profiles is 2.0\%, 2.52\%, and 19.1\% for A611, A963, and $M_{gal} = 10^{13}M_{\odot}$ galaxies bin, respectively. It is likely that the percentages are related to the location of the Gaussian peak as the fit modification for the Chae et al. galaxies' profile was different than the other fits.  This is because its Gaussian peak was located at $\epsilon=0.34\phi_0$ whereas the others are closer to $\epsilon=0.85\phi_0$. 

To check if the Gaussian shape provides a reasonable approximation to the difference between the energy distributions of DARKexp and the data, we calculated the shape of the actual difference between $N(E)$ derived from galaxy cluster A611 density profile and DARKexp $N(\epsilon)$ with $\phi_0=3.5$. The result is a skewed peak shown in Figure~\ref{Diff peak}. (We calculated the energy distribution of A611 from its density profile with an algorithm outlined by \citet{Bin82}.)  The peak is similar enough to a Gaussian to warrant the use of equation~\eqref{Nmod}.

We emphasize that the modifications added to DARKexp $N(\epsilon)$ to reproduce simulated halos and observed systems are the same, both given by equation~\eqref{Nmod}. This suggests that both types of systems share the same characteristic, which we argue is their state of incomplete relaxation in the central regions. The fact that the amplitude of modification is smaller in the case of simulated halos (top panels of Figure~\ref{6panel}) than observed systems (Figure~\ref{Observation Nplots}) argues that the former are closer to being fully relaxed that the latter. 

We note that the modifications we added to DARKexp in this Section are consistent with our findings in Section~\ref{class1}: making the $N(\epsilon)$ slope less negative around $\epsilon=0.8\phi_0$ results in greater density slope oscillations in the corresponding density profiles. 

We also note that there may be some degeneracy between the Gaussian parameters $a$, $\bar{\epsilon}$ and $\sigma$ (equation \eqref{Nmod}) and the starting central potential $\phi_0$, when fitting the density profiles. Solutions may 
not be unique as one set of parameters with a central potential can give similar results to another central potential with different modification parameters. 

\section{Conclusions}\label{Conclusions}

As we show in this paper, many, if not most of these systems exhibit non-monotonic changes in slope, or 
oscillations. Galaxies and clusters, whether observed or simulated can be described collectively as dynamically evolved because they are the result of 
evolution, either in the real universe, or in a computer.  Monotonically changing density profile slopes appear to be the exception rather than the rule in the very central regions of observed and simulated galaxies 
and clusters, and pure dark matter halos. 

Oscillations are also common in other types of systems, those that were arrived at starting from a distribution function of a differential energy 
distribution, such as the isothermal sphere and DARKexp.  Systems that do not have slope oscillations are mostly those that were constructed
that way: NFW, Einasto, Hernquist, etc.

In this paper, we study the two classes of systems whose profile slopes do oscillate: systems based on a physical principle or derivation (Class 1), and 
expressed in terms of $f(E)$ or $N(E)$; we call these physics-based, and dynamically evolved systems (Class 2). The two types of oscillations appear to be 
unrelated, and can be superimposed in a single system. Slope oscillations in the dynamically evolved systems occur at $-2\lesssim\log(r/r_{-2})\lesssim 0$, while those in the relaxed, collisionless, physics-based systems (DARKexp) start at smaller radii, $\log(r/r_{-2})\lesssim -2$.

We argue that slope oscillations in the dynamically evolved systems occur because these systems are not fully relaxed. To support our claim, we show that 
adding a small perturbation to the $N(E)$ of a fully relaxed system, namely DARKexp, reproduces the oscillating density profiles of observed and simulated 
systems. The fact that the modifications to $N(E)$ are small and affect a small fraction of the systems' energy distribution implies that these systems are close to being fully 
relaxed. 

Non-monotonically changing density slopes (or oscillations) in physics-based systems have a different origin. It appears that a generic monotonically changing 
$f(E)$ or $N(E)$ does not produce a monotonically changing density slope. This conclusion is based on the behavior of the systems like isothermal spheres, 
King profiles, certain polytropes, and DARKexp, as well as our (limited) exploration of other functional forms for $N(E)$. We are not claiming that all 
physics-based systems show density slope oscillation, but only that these are common. Focusing on the density slope oscillations at small radii, we then ask 
if the magnitude of these oscillations can be predicted solely from the shape of the corresponding $N(E)$. We devise a metric to quantify density slope 
oscillations, and show that it is related to a property of $N(E)$, namely $d\log(N)/d\epsilon$ measured at $\epsilon=0.8\phi_0$. While the relation between 
the two has some scatter, it does confirm our claim that density slope oscillations are the consequence of the shape of $N(E)$.

The most interesting avenue for future work is to further investigate the dynamically evolved systems (Class 2). We plan to investigate the time evolution of simulated systems with the goal of measuring how $N(E)$ evolves over time and what corresponding effects this has on the density profile for halos. For the simulated systems, where everything is known about the constituent particles, it is possible to obtain the system's actual $N(E)$, instead of calculating it from the density profile, as we have done here. If the deviation of the $N(E)$ from that of the relaxed DARKexp is indeed a small deficit of particles, localized in energy, then it would be interesting to investigate why the system did not attain full relaxation. Was it that the particles were not able to freely exchange angular momentum during evolution, as we speculate, or something else?  If it is, for example, angular momentum, then the similarity between density slope oscillations in simulated and observed systems (see Fig. \ref{6panel} and \ref{Observation Fits}) suggests that the two types of systems have similarities in their distributions of angular momentum. This information can give us a new tool to understand the dynamics within the observed galaxies and clusters, and add to the current tool kit which consists of mass distribution obtained from lensing and line of sight velocity dispersion(s), or velocity distributions. 

More generally, by comparing a system's properties, and especially $N(E)$, to 
those of DARKexp, one can better ascertain a system's dynamical state.  Recall the hierarchy of states that a self-gravitating system can find itself in. 
Fully relaxed is not the same as virialized, or being in Jeans (hydrostatic) equilibrium. A virialized system possesses some global, or integral properties, 
but need not satisfy anything else. Jeans equilibrium implies more than just the virial equilibrium, but it does not, strictly speaking, guarantee that the 
system is even stable. Most galaxies and clusters, real and simulated, that have a smooth circular/spherical or elliptical appearance over some prolonged 
period of time are likely in stable Jeans equilibrium. But even that does not mean that they are fully relaxed. The state of full relaxation means that the 
system is not only in a long term stable equilibrium, but that it has also erased its formation and evolution history. While determining if a system is in 
stable Jeans equilibrium is relatively easy (in simulations and observations), it is difficult to know if it is relaxed. Comparison to DARKexp, a theoretically 
derived collisionlessly relaxed model, might be the best way to establish that.

\acknowledgments

We thank Andrew Newman for sharing their data on galaxy clusters, and Andrew Newman
and Kyu-Hyun Chae for answering our questions about their work over email.

\end{document}